# ФЕРРОЭЛЕКТРИЧЕСКИЙ СПИН-ВОЛНОВОЙ РЕЗОНАНС

ЕРЧАК Д.П [a], ЕРЧАК Е.Д [b], КИРИЛЕНКО А.И [a], ПОПЕЧИЦ В.И [a,b]

[a] *Минский государственный высший авиационный колледж, Минск*
[b] *Белорусский государственный университет, Минск*

**Введение**

Для периодической цепочки, состоящей из *n* двухуровневых подсистем, взаимодействующих между собой и с внешним осциллирующим полем, на основе метода операторов перехода в [1] были получена система операторных разностно-дифференциальных уравнений, переходящая в систему операторных дифференциальных уравнений в континуальном пределе. Данная система была сведена в одно векторное операторное уравнение. Полученное уравнение оказалось операторным эквивалентом известного уравнения Ландау- Лифшица (Л-Л) [2], введенному изначально для макроскопического описания движения вектора намагниченности в ферромагнетиках. В результате, с квантово-механических позиций было строго обоснована применимость (Л-Л)-уравнения в спектроскопии как базового уравнения динамики спектроскопических переходов. Было установлено также, что с помощью данного уравнения математически и физически с единых позиций может быть исследована эволюция динамических систем при магнитном и оптическом резонансах. В результате вместо оптических уравнений Блоха были предложены новые уравнения динамики оптических переходов. В них вместо вектора Блоха, имеющего физически разнородные компоненты, одна из которых вообще не связана с электрическими характеристиками системы, фигурирует вполне конкретная физическая векторная величина - электрический дипольный момент. В частности, указанная величина есть электрический "спиновый" момент, то есть электрический собственный момент исследуемых частиц или квазичастиц, связанный с наличием у них спина. Заметим, что электрический собственный момент впервые был предсказан математически Дираком в его знаменитой работе [3]. Вторая векторная величина, имеющая в оптических уравнениях Блоха также физически разнородные компоненты, в предложенных уравнениях есть вектор электрического поля. Математическая идентичность модели для магнитных и электрических дипольных переходов приводила к выводу, что ряд физических явлений, сопровождающих магниторезонансные и оптические переходы, должен быть качественно одинаков. Как результат, в [1] были предсказаны электрические спин-волновые резонансы, являющиеся аналогами известных магнитных спин-волновых резонансов. Данные резонансы были действительно обнаружены, причем как ферроэлектрический спин-волновой резонанс (ФЭ СВР) [4], так и антиферроэлектрический спин-волновой резонанс (АФ ЭСВР) [5]. Их экспериментальное наблюдение явилось прямым подтверждением корректности теоретической модели, предложенной в [1].

В настоящей работе изложены детально результаты целенаправленного эксперимента для идентификации первого оптического аналога магнитного спин-волнового резонанса, - ферроэлектрического спин-волнового резонанса, и проведено сравнение с известными литературными данными, которые были реинтерпретированы.

**Методика и техника эксперимента**

Наиболее подходящими для идентификации ФЭ СВР нам представлялись образцы карбиноидов, являющиеся системами углеродных квазиодномерных цепочек. Было проведено исследование набора из 20 образцов карбиноидов, полученных путем химической дегидрогалогенизации пленок поливинилиденфторида (ПВДФ). Весь набор состоял из 5 групп, отличающихся параметрами процесса дегидрогалогенизации и последующей обработки. В данной работе сообщаются результаты исследования инфракрасного (ИК) отражения, ИК-поглощения и комбинационного рассеяния (КР) только для двух на наш взгляд наиболее интересных групп образцов $A$ и $B$ (в дальнейшем $A$-образцы и $B$-образцы) с одноосно ориентированными цепочками. К тому же данные образцы были исследованы ранее методом электронного парамагнитного резонанса (ЭПР) [6, 7], что позволяет сравнить результаты оптических и ЭПР исследований. Образцы хранились после их изготовления ~ 12 лет при комнатной температуре. Исследования ИК-отражения, ИК-поглощения были проведены в диапазоне 400 – 5000 см$^{-1}$ на ИК-спектрометре "Nexus" с Фурье преобразованием. Угол падения света в исследованиях ИК-отражения составлял 20°. КР-спектры регистрировались в 180°-геометрии с помощью монохроматора "Spectra Pro 500i" фирмы Acton Research Corporation. Мощность излучения лазера с длиной волны 532 нм не превышала значения 30 мВт.

**Экспериментальные результаты**

Согласно [6 - 9], в карбиноидах, подобно *транс*-полиацетилену (*т*-ПА), основными носителями заряда и спина являются топологические солитоны, названные спин-Пайерлс солитонами. Спин-Пайерлс солитоны, как и топологические солитоны в *т*-ПА, являются активными в электронно-колебательных спектрах. В частности, наборы линий $\{a, b, c\}_{[A]}$ с максимумами при частотах $\{\nu_a\ \nu_b\ \nu_c\}_{[A]} = \{1730, 1630, 1068\}$ см$^{-1}$ в спектрах ИК-поглощения $A$-образцов и $\{a, b, c\}_{[B]}$ с максимумами при частотах $\{\nu_a\ \nu_b\ \nu_c\}_{[B]} = \{1750, 1650, 1080\}$ см$^{-1}$ в спектрах ИК-поглощения $B$-образцов были идентифицированы как локализованные колебательные моды спин-Пайерлс солитонов [8, 9]. Следует заметить, что данные линии из-за сильного поглощения могли быть зарегистрированы только для молотых образцов, спрессованных в таблетку с *KBr*. При этом, естественно, терялась информация, обусловленная трансляционной симметрией исходных образцов. Изучение оптического отражения вместо поглощения (пропускания) позволило избежать отмеченной трудности. Поскольку, как было установлено, отражение носит объемный характер (а не зеркальный поверхностный), анализ спектров отражения существенно упростился. Было также установлено, что различие между спектрами образцов, принадлежащих одной и той же группе, не существенно, и далее будут представлены и проанализированы зкспериментальные данные для двух образцов *A* и *B* (произвольно выбранным из *A*- и *B*–групп, соответственно).

ИК-спектры отражения *A*- и *B*-образцов представлены на рис.1 и рис.2. Общий вид спектрального распределения интенсивности сигнала отражения в области 400 – 5000 см$^{-1}$ представлен на рис.1*а* и рис.2*а* для *A*- и *B*-образцов, соответственно. Детальная структура спектров тех же образцов в области 800 - 2200 см$^{-1}$ представлена, соответственно, на рис.1*б* и рис.2*б*.

Было обнаружено, что две линии *a*, *b* из набора {*a*, *b*, *c*} локализованных колебательных мод спин-Пайерлс солитонов, идентифицированных в молотых образцах [8, 9], расщепляются на ряд компонент, представляющих наборы {$a_n$} и {$b_n$}. В *A*-образце надежно зарегистрированы были 3 компоненты с частотами {$\nu_n(a)$}$^3_{[A]}$ = {1750.5, 1473.6, 904.6}(± 0.2) см$^{-1}$ в {$a_n$}$_{[A]}$-наборе, и также 3 компоненты с частотами {$\nu_n(b)$}$^3_{[A]}$ = {1664.4, 1407.4, 889.2}(± 0.2) cm$^{-1}$ в наборе {$b_n$}$_{[A]}$, рис.1. Аналогично, в *B*-образце {$a_n$}$_{[B]}$-набор состоит из линий, спектральное положение которых задается частотами из набора {$\nu_n(a)$}$^3_{[B]}$ = {1738.3, 1464, 903.6} (± 0.2) см$^{-1}$. Набор {$b_n$}$_{[B]}$ в *B*-образце состоит из линий в положении при {$\nu_n(b)$}$^3_{[B]}$ = {1685.7, 1419.4, 886.7} (± 2) см$^{-1}$, рис.2. Calculated frequency position values for uneven MSWROA - modes in B-sample are: $\nu_1$ = 1738.3 cm$^{-1}$, $\nu_3$ = 1460 cm$^{-1}$, $\nu_5$ = 903.6 cm$^{-1}$ in a$_{[B]}$ -set and $\nu_1$ = 1685.7, $\nu_3$ = 1419.3, $\nu_5$ = 886.7 cm$^{-1}$ in b$_{[B]}$ –set.Третья солитонная мода *c* наблюдается при несколько иных значениях частоты внешнего воздействия – при 1088.4 cm$^{-1}$ в *A*-образце и при 1074.4 см$^{-1}$ в *B*-образце (пик моды *c* находился при 1068 см$^{-1}$ и при 1080 см$^{-1}$ в молотых *A*- и *B*-образцах, соответственно). Ее относительная интенсивность в обоих ориентированных образцах заметно меньше в сравнении с интенсивностью линии *c* в молотых образцах. Расщепление моды *c* в обоих ориентированных образцах не было зарегистрировано. Существенным моментом является то, что положения отдельных компонент в наборах {$a_n$} и {$b_n$} в обоих образцах могут быть заданы феноменологически законом, идентичным в своей математической форме закону Киттеля для ферромагнитного спин-волнового резонанса (ФМ СВР), полученного для ферромагнитно упорядоченной линейной цепочки [10]:

$$\nu_n = \nu_0 - \mathcal{A} n^2, \qquad (1)$$

где $n \in \mathrm{N}$, включая нуль, $\nu_n$ - частота моды с номером $n$, $\mathcal{A}$ - параметр материала ($\mathcal{A} > 0$). Расчетные значения частот мод, найденные для $A$-образца согласно (1), есть $\{\nu_n(a)\}^{\mathrm{T}}_{[A]}$ = $\{1785.7, 1750.5, 1644.7, 1468.5, 1221.8, 904.6\}$ cm$^{-1}$ для набора $\{a_n\}$, $n = \overline{0,5}$, и $\{\nu_n(b)\}^{\mathrm{T}}_{[A]}$ = $\{1698, 1664.4, 1568.6, 1406.8, 1180.4, 889.2\}$ cm$^{-1}$ для набора $\{b_n\}$, $n = \overline{0,5}$. Как следует из сопоставления расчетных и экспериментальных значений частот мод и непосредственно видно из рис.1б, на котором положение расчетных нечетных компонент обозначено двумя рядами стрелок - верхний ряд соответствует набору $\{a_n\}$, нижний ряд соответствует набору $\{b_n\}$, наблюдаются лишь нечетные моды, причем налицо хорошее согласие экспериментальных значений частот нечетных компонент в наборах $\{a_n\}$ и $\{b_n\}$ с вычисленными. Моды, соответствующие четным $n$, не наблюдаются. Совершенно аналогичная картина имеет место и в спектре $B$-образца. Как и в $A$-образце $\{a_n\}_{[B]}$- и $\{b_n\}_{[B]}$- наборы спектральных линий представляют собой моды спин-волнового резонанса с нечетными номерами 1, 3, 5. Расчетные положение данных мод в спектре, указанное на рис.2б стрелками аналогично рис.1б, соответствует следующим значениям частот $\{\nu_n(a)\}^{\mathrm{T}}_{[A]}$ = $\{1738.3, 1460, 903.6$ cm$^{-1}\}$, $\{\nu_n(b)\}^{\mathrm{T}}_{[B]}$ = $\{1685.7, 1419.3, 886.7$ cm$^{-1}\}$. Видим, что согласие вычисленных и экспериментальных значений частот мод также хорошее. Амплитуда мод, как видно из рис.1, рис.2, уменьшается с ростом номера моды. Экспериментальный вид данной зависимости был установлен для образца $A$ – амплитуда моды обратно пропорциональна номеру моды, рис.3.

Выполнение закона дисперсии, совпадающего с законом Киттеля, означает, что на исследуемых образцах наблюдается оптический аналог ферромагнитного спин-волнового резонанса, соответствующий фероэлектрическому упорядочению электрических дипольных моментов, то есть фероэлектрический спин-волновой резонанс. Заметим, что совершенно аналогичная картина, то есть преимущественное возбуждение нечетных резонансных мод, имела место при первом экспериментальном наблюдении ФМ СВР в пермаллоевых ферромагнитных пленках [11]. Существенно также, что ФМ СВР, сопровождавшийся возбуждением также только нечетных резонансных мод, наблюдался и в исследуемых нами образцах карбиноидов, детальное изложение которого приведено в [6, 7]. Таким образом, мы можем утверждать об открытии нового физического явления – аналога ферромагнитного спин-волнового резонанса – фероэлектрического спин-волнового резонанса.

Значения параметра $\mathcal{A}$ в соотношении (1), соответственно, составляют $\mathcal{A}_{[A]}(a)$ = 35.2 см$^{-1}$ и $\mathcal{A}_{[A]}(b)$ = 33.6 см$^{-1}$ для наборов $\{a_n\}$ и $\{b_n\}$ в ИК-спектрах *A*-образца и $\mathcal{A}_{[B]}(a)$ = 34.7 см$^{-1}$ и $\mathcal{A}_{[B]}(b)$ = 33.2 см$^{-1}$ для наборов $\{a_n\}$ и $\{b_n\}$ в ИК-спектрах *B*-образца. Несколько меньшие значения параметров $\mathcal{A}_{[B]}$ в *B*-образцах в сравнении с параметрами $\mathcal{A}_{[A]}$ в *A*-образцах свидетельствуют о меньшей величине межсолитонного взаимодействия в *B*-образцах, что хорошо согласуется с результатами предыдущих исследований в [8, 9].

Сам факт наблюдения ФМ СВР и ФЭ СВР в одних и тех же образцах серии *A* на одних и тех же объектах – спин-Пайерлс солитонах – означает, что спин-Пайерлс солитоны обладают, как магнитным, так и электрическим моментом, причем в образцах указанной серии имеет место как ферромагнитное, так и ферроэлектрическое упорядочение. Таким образом, карбиноидные *A*–образцы являются одновременно ферромагнетиками и ферроэлектриками. Учитывая результаты работы [5], следует дополнить, что они же являются и антиферроэлектриками. В то же время карбиноидные *B*–образцы являются только ферроэлектриками и антиферроэлектриками, поскольку ФМ СВР на них не наблюдался. Заметим, что карбиноиды являются (насколько нам известно) единственной полимерной (и органической вообще) системой, обладающей одновременно свойствами электрического и магнитного упорядочения. Для неорганических веществ возможность одновременного проявления свойств электрического и магнитного упорядочения является хорошо известной. Например, кристаллы $BaMnF_4$ проявляют одновременно ферроэлектрические, антиферроэлектрические и антиферромагнитные свойства [12]. Заметим также, что мы имеем дело с ферромагнетизмом и сегнетоэлектричеством нового вида. Природа возникновения антиферроферроэлектрических свойств в изучаемых карбиноидах детально изложена в [5]. Попутно в [5] был также упомянут механизм возникновения ферроэлектрических и ферромагнитных свойств. Принципиальным отличием как ферромагнитного, так и электрического упорядочения обоих видов в карбиноидах от известных явлений ферромагнетизма, ферроэлектричества и антиферроэлектричества является то, что в атомных цепочках карбиноидов имеет место упорядочение носителей спина, локализованных в валентных связях, то есть, в конечном итоге, в *наружных* оболочках атомов углерода. В то же время хорошо известные ферромагнетизм, ферроэлектричество и антиферроэлектричество определяются незаполненными *внутренними* оболочками атомов переходных металлов или редкоземельных элементов.

Исследуемые образцы являются также КР-активными. Спектры КР, регистрируемые в них, характеризуются рядом относительно узких компонент, перекрывающихся с очень широкой фоновой линией, рис.2в, рис.2г. Фоновая линия достигает максимального значения при ~ 2429 см$^{-1}$ и ~2388 см$^{-1}$ в спектрах *A*- и *B*-образцов соответственно. Естественно предположить, что присутствие фоновой линии есть свидетельство наличия квазинепрерывного смещения частоты рассеянных фотонов, и, следовательно, может указывать на расщепление характеристической (то есть имеющей частоту в пределах ~ 2000 - 2500 см$^{-1}$ [13]) для карбинов и карбиноидов моды колебаний в энергетическую зону с эффективной шириной порядка указанных максимальных значений. В пользу данного предположения свидетельствует большее значение частоты в максимуме спектрального распределения интенсивности КР, а, следовательно, и ширины зоны в *A*-образце в сравнении с *B*-образцом. В свою очередь, большее значение ширины зоны в *A*-образце говорит о большей величине взаимодействия между локализованными центрами, ответственными за данную колебательную моду, то есть между спин-Пайерлс солитонами. Действительно, данный вывод находится в полном согласии с тем фактом, что в *A*-образцах имеет место ферромагнитное упорядочение, в то время как в *B*-образцах оно отсутствует. Более того, значения $\mathcal{A}_{[A]}(a) = 35.2$ см$^{-1}$ и $\mathcal{A}_{[A]}(b) = 33.6$ см$^{-1}$ для наборов $\{a_n\}$ и $\{b_n\}$ в ИК-спектрах *A*-образца и $\mathcal{A}_{[B]}(a) = 34.7$ см$^{-1}$ и $\mathcal{A}_{[B]}(b) = 33.2$ см$^{-1}$ для наборов $\{a_n\}$ и $\{b_n\}$ в ИК-спектрах *B*-образца конкретно указывают, насколько величина солитон-солитонного взаимодействия в *A*-образцах больше данной величины в *B*-образцах.

Положение относительно узких линий в спектрах КР, подобно наборам $\{a_n\}$ и $\{b_n\}$ в спектрах ИК-отражения, также может быть задано соотношением (1). Действительно, например, в *B*-образце зарегистрирован набор из трех компонент, который обозначим $\{a_{Rn}\}$, с частотами $\{\nu_n(a_R)\}^{э}_{[B]} = \{1605, 1364, 1064\}$ (± 2 см$^{-1}$). Положение вычисленных согласно (1) значений частот мод с номерами 0, 1, 2, 3 составляет соответственно $\{\nu_n(a_R)\}^{т}_{[B]} = \{1604, 1544, 1364, 1064\}$ cm$^{-1}$. Как видно из рис.2г, на котором положения вычисленных мод обозначены стрелками, расчетные значения частот мод с номерами 0, 2, 3 хорошо согласуются с экспериментальными значениями частот пиков из набора $\{\nu_n(a_R)\}^{э}_{[B]}$. Положение моды 1 непосредственно не регистрируется. Однако асимметрия контура нулевой моды в интервале (1500 – 1650) см$^{-1}$, косвенно указывает на ее присутствие и перекрытие с нулевой модой в данном спектральном интервале. Основанием для данного утверждения является то, что, например, вторая мода достаточно симметрична, а функция формы всех мод должна быть одной и той же (см.далее соотношение (10)). Значение частоты в максимуме данной моды было получено путем декомпозиции спектра на участке 1500 – 1650 см$^{-1}$ на 2 компоненты и составило 1545 см$^{-1}$ (± 4 см$^{-1}$) в согласии с выше приведенными расчетными данными. Детальный анализ при большем усилении других участков спектра рис.2в показал присутствие еще ряда компонент. Среди них линии с максимумами при 643 см$^{-1}$ и 105 см$^{-1}$ (± 2 см$^{-1}$) следует отнести к спин-волновым модам вышеприведенного набора с номерами 4 и 5 соответственно. Положение 4-й, 5-й мод, вычисленное согласно (1), составляет, соответственно, 644 см$^{-1}$ и 104 см$^{-1}$, то есть хорошо согласуется с наблюдаемыми значениями.

Таким образом, в спектре ФЭ СВР, детектируемом с помощью КР, пять мод из набора {$a_{Rn}$} (0-я, 2-я, 3-я, 4-я, 5-я) зарегистрированы напрямую (по положению их пиков), присутствие 1-й моды установлено косвенно из изучения формы контура спектра КР в области (1500 – 1650) см$^{-1}$. Значение параметра спин-волнового расщепления $\mathcal{A}_{[В]}(a_R)$, определяющее указанный набор в КР-(ФЭ СВР)-спектре, оказалось равным 60 см$^{-1}$.

В спектрах КР имеется и второй набор, обозначим его {$b_{Rn}$}, мод ФЭ СВР, который, несмотря на неблагоприятное соотношение сигнал/шум, все же можно идентифицировать. Это линии с частотами из набора {$\nu_n(b_R)$}$^э_{[B]}$ = {1511, 1269, 958, 515} (± 12) см$^{-1}$. Вычисленные согласно (1) первые пять значений частот $\nu_n(b_R)$, $n = \overline{0,4}$, второго набора есть {$\nu_n(b_R)$}$^т_{[B]}$ = {1517.8, 1455.6, 1269, 958, 522} cm$^{-1}$ Из сравнения {$\nu_n(b_R)$}$^т_{[B]}$ с экспериментальным набором {$\nu_n(b_R)$}$^э_{[B]}$ видно, что в наборе {$b_{Rn}$} экспериментально наблюдаются нулевая, вторая, третья и четвертая моды. Экспериментальное значение частоты первой моды, перекрывающейся с нулевой модой, установлено не было. Значение параметра расщепления $\mathcal{A}_{[В]}(b_R)$, определяющее второй набор в спектре КР-(ФЭ СВР), следующее $\mathcal{A}_{[В]}(b_R)$ = 62.2 см$^{-1}$. Налицо два основных отличия спектров ФЭ СВР, регистрируемых методом КР и ИК. Спектры ФЭ СВР, регистрируемые методом ИК, содержат только нечетные моды. В спектрах КР-(ФЭ СВР), наблюдаются как четные, так и нечетные моды. Еще более существенным является возрастание параметра $\mathcal{A}$ практически в 2 раза (в пересчете на одинаковое значение частоты нулевых мод в линейной аппроксимации) в сравнении со значением $\mathcal{A}$ из данных ИК-отражения. Данный результат соответствует предсказанию работы [1] и является независимым аргументом для интерпретации обнаруженного явления расщепления колебательных термов как ФЭ СВР.

В работе [1], как упоминалось, было получено квантовомеханическое уравнение динамики квантовых переходов для периодической цепочки, состоящей из $n$ двухуровневых подсистем, взаимодействующих между собой и с внешним осциллирующим полем. Взаимодействие между элементами цепочки было задано гамильтонианом квантовой XYZ – модели Гейзенберга в случае магнитных дипольных моментов и соответствующим ему аналогом в случае электрических дипольных моментов. В континуальном пределе модифицированное уравнение для наблюдаемых, описывающее динамику оптических переходов с учетом релаксации, имеет вид:

$$\frac{\partial \vec{S}(z)}{\partial t} = \left[\vec{S}(z) \times \gamma_E \vec{E}\right] - \frac{4a^2 J}{\hbar^2}\left[\vec{S}(z) \times \nabla^2 \vec{S}(z)\right] + \frac{\vec{S}(z) - \vec{S}_0(z)}{\tau} \qquad (2)$$

где $\vec{S}(z)$ является электрическим аналогом "спинового" магнитного момента, $\gamma_E$ – гироэлектрическое отношение, являющееся оптическим аналогом гиромагнитного отношения, $J_E$ есть оптический аналог константы обменного взаимодействия, $a$ – период решетки цепочки, $\hbar$ – постоянная Планка, $\vec{E}$ – электрическое поле, $\vec{S}_0(z)$ – равновесное значение электрического "спинового" момента в точке $z$, направление которого совпадает с направлением оси $z$, $\tau$ – время релаксации. Линеаризация уравнения (4) достигается, если принять во внимание, что экспериментальные значения амплитуд компонент внешнего осциллирующего поля $E^x$, $E^y$, входящие в (4), значительно меньше величины $z$-компоненты внутрикристаллического электрического поля $E^z$. Аналогичное соотношение имеет место для компонент результирующего электрического "спинового" момента. Таким образом, имеем:

$$E^x, \ E^y \ll E^z, \ S^x(z), \ S^y(z) \ll S^z(z). \tag{3}$$

Стационарное распределение $\vec{S}_0(z)$ спинового момента вдоль цепочки считаем однородным, тогда, как легко показать, линеаризованные уравнения для компонент вектора $\vec{S}(z)$ будут иметь вид:

$$\frac{\partial S^+(z)}{\partial t} = iDS \frac{\partial^2 S^+(z)}{\partial z^2} - i\omega_0 S^+(z) + i\gamma_E S E^+ + \frac{S^+(z)}{\tau}, \tag{4}$$

где

$$S^+(z) = S^x(z) + iS^y(z). \tag{5}$$

Будем искать решение уравнения (4) в виде:

$$S^+(z) = \exp(-i\omega t) \sum_{n=-\infty}^{n=\infty} a_n \exp(ik_n z) \tag{6}$$

Подставляя (6) в (4), получаем соотношение

$$\sum_{n=-\infty}^{n=\infty} \left\{ a_n \exp(ik_n z) \left( DS k_n^2 + \omega_0 - \omega + \frac{i}{\tau} \right) \right\} = \gamma_E S E_1. \tag{7}$$

В качестве граничных условий выберем периодические граничные условия:

$$S^+(z) = S^+(z+L), \ \exp(ikL) = \pm 1, \tag{8}$$

где $L$ – длина цепочки. Из граничных условий сразу определяются допустимые значения величины квазиволнового вектора $\pm|\vec{k}|$: $k_n = \frac{n\pi}{L}$, где $n = 0, \pm 1, \pm 2, \pm \ ...$ Тогда из (7) получаем закон дисперсии, а также соотношение для зависимости функции формы и значений амплитуды резонансных мод от номера моды:

$$\nu_n = \nu_0 + DSk_n^2, \quad D = \frac{2Ja^2}{\pi\hbar^2}, \quad \nu_0 = \frac{\gamma_E E^z}{2\pi}, \quad a_n = \begin{cases} -\dfrac{i\gamma_E S\tau^2 E_1}{\pi n} \dfrac{\left[(\omega_n - \omega) - \dfrac{i}{\tau}\right]}{\left[1 + (\omega_n - \omega)^2 \tau^2\right]}, & n = 1, 3, 5, ... \\ 0, & n = 2, 4, 6, ... \end{cases} \quad (9)$$

В соотношениях (9) $S$ – значение спина. Видно, что соотношения для закона дисперсии в (9) совпадают с экспериментально установленными соотношениями (1). Действительно, для этого достаточно положить $\mathcal{A} = -DS$, что возможно, поскольку $J < 0$ при ферроэлектрическом упорядочении, то есть $J = -|J|$. Как обычно, $Re\, a_n$ соответствует сигналу поглощения, $Im\, a_n$ сигналу дисперсии. Заметим, что возбуждение только нечетных мод есть следствие однородности поля вдоль цепочки. Условие однородности поля заведомо выполнялось в измерениях ИК-отражения и ИК-поглощения. Это условие, однако, не выполнялось в измерениях КР. В случае неоднородного внешнего возбуждения правая часть в (7) становится зависимой от $z$. В результате получаем значения $a_n$, отличные от нуля для всех $n$. Другими словами, тогда становится возможным наблюдение и четных мод, что действительно имеет место при исследовании КР, рис.2. Как следует из (9), форма мод ФЭ СВР в рамках предложенной модели является лоренцевой. Из (9) следует также, что амплитуда моды обратно пропорциональна номеру моды. Данная зависимость представлена на рис.3 сплошной кривой. Как видно из рисунка, экспериментальные данные ИК-отражения хорошо согласуются с предсказываемой зависимостью. В то же время сравнительно высокая плотность энергии лазерного излучения, используемая при изучении КР, может приводить к нарушению соотношений (3) и, следовательно, также к иному соотношению между амплитудами резонансных мод, даже если закон дисперсии сохраняется в виде (9).

Имеющиеся данные для параметра расщепления $\mathcal{A}^H$ в спектрах МСВР, составляющие для $A$- образца $\mathcal{A}_a^H = 23.7$ Гс и $\mathcal{A}_b^H = 28.9$ Гс [6], позволяют получить величину отношения обменных констант $J_E/J_H$. Данная величина равна ~1.6 $10^4$ для $\{a_n\}_{[A]}$–наборов и ~1.2 $10^4$ для $\{b_n\}_{[A]}$ –наборов наблюдаемых обоими методами спин-волновых компонент. Учитывая соотношение, связывающее обменное взаимодействие с зарядом и пренебрегая изменением координатного распределения волновой функции спин-Пайерлс солитонов при наличии внешнего постоянного магнитного поля и в его отсутствие, получаем оценку для отношения компонент комплексного заряда спин-Пайерлс солитонов $\dfrac{g}{e} \approx \sqrt{\dfrac{J_E}{J_H}} \approx (1.1 - 1.3)10^2$. Видно, что данный результат хорошо согласуется с соотношением, известным из теории квантования заряда Дирака [14], g ≈ 68.5 e n, где n = ±1, ±2,..., при n = 2. Некоторое отклонение вполне допустимо из-за грубости оценки.

Следует заметить, что экспериментальные данные, указывающие на явление ФЭ СВР содержатся в [15], хотя само явление идентифицировано в [15] не было. Winter и Kuzmany исследовали монокристаллы калийсодержащего фуллерида методом КР при 80K [15]. Было обнаружено, что две низкочастотные колебательные $Hg$-моды фуллерена претерпевают расщепление, отсутствующее в нелегированных монокристаллах. Участок спектра, соответствующий $Hg(2)$-моде, воспроизведенный из [15], представлен на рис.4а. Как видно из рисунка, наблюдаемая структура является типичной для ФЭ СВР. В то же время Winter и Kuzmany связывают происхождение расщепления со снятием пятикратного вырождения $Hg$-мод, производя декомпозицию наблюдаемого спектра на пять компонент, рис.4а. На самом деле в спектре четко выражены только 4 компоненты со значениями частот 425 $cm^{-1}$, 413 $cm^{-1}$, 388 $cm^{-1}$, 351 $cm^{-1}$. Если же принимать во внимание наличие пятой компоненты, соответствующей плечу при ~ 400 $cm^{-1}$, тогда следует принять во внимание наличие шестой и седьмой компонент, соответствующих плечу при ~ 370 $cm^{-1}$ и пику (хотя и слабо выраженному) при ~ 300 $cm^{-1}$, более четко видным из рис.5b в [15] (рис.5b здесь не воспроизводится), поскольку их интенсивность вполне сравнима с интенсивностью компоненты при ~ 400 $cm^{-1}$. Gunnarsson [16] заметил, что предложенная Winter and Kuzmany модель никак не объясняет существенную зависимость ширины моды от ее номера. Более того в рамках модели Winter and Kuzmany указанной зависимости не должно быть. Модель Winter and Kuzmany не позволяет также получить теоретические резонансные значения мод в спектре.

Идентификация данного явления как ФЭ СВР сразу же объясняет позиции наблюдаемых мод в спектре в соответствии с законом дисперсии (1). Действительно, вычисленные согласно (1) резонансные значения для десяти мод следующие $\nu_0$ = 426.6 $cm^{-1}$, $\nu_1$ = 425 $cm^{-1}$, $\nu_2$ = 420.4 $cm^{-1}$, $\nu_3$ = 412.7 $cm^{-1}$, $\nu_4$ = 401.9 $cm^{-1}$, $\nu_5$ = 388 $cm^{-1}$, $\nu_6$ = 371 $cm^{-1}$, $\nu_7$ = 351 $cm^{-1}$, $\nu_8$ = 327.9 $cm^{-1}$, $\nu_9$ = 301.7 $cm^{-1}$, параметр $\mathcal{A}$ = 1.6 $cm^{-1}$. Расчетные положения мод обозначены на рис.4b стрелками. Как видно из рисунка, хорошо наблюдаемые 4 компоненты соответствуют четырем нечетным модам ФЭ СВР $\nu_1$, $\nu_3$, $\nu_5$, $\nu_7$ Преимущественное возбуждение нечетным мод в соответствии с вышеизложенной линеаризованной моделью свидетельствует о достаточно однородном возбуждении образца с низким уровнем, удовлетворяющем условиям (5). Плечо при ~ 400 $cm^{-1}$, рис.4, а также плечо при ~ 370 $cm^{-1}$, рис.5b в [15], соответствуют четным модам $\nu_4$ и $\nu_6$, что указывает на некоторую неоднородность внешнего воздействия, однако весьма незначительную, поскольку моды $\nu_2$ и $\nu_8$ маскируются в спектре, в то время как даже пятая нечетная мода $\nu_9$ регистрируется достаточно уверенно, рис.5b в [15]. Далее, очевидно, что моды с n = 0 и n = 1 из-за малой величины $\mathcal{A}$ СВР-расщепления дают одну компоненту в спектре СВР. Возрастание же ширины моды с ростом ее номера является характерным для ФМ СВР, см., например, [17] и связывается с n − кратным увеличением числа длин волн, укладывающихся на единице длины кристалла, то есть n − кратным увеличением числа элементарных возбуждений для моды с номером n в сравнении с модой n = 1. В результате мы имеем n−кратное усиление спин-фононного взаимодействия (в предположении равномерного распределения фононов в данном спектральном интервале), приводящего к соответствующему уширению. Очевидно, предположение равномерного распределения фононов хорошо выполняется в ФМ СВР из-за малых в абсолютном отношении изменений энергии при переходе от одной моды ФМ СВР к другой. Достаточно хорошо оно выполняется и при малых (~1 $cm^{-1}$) величинах $\mathcal{A}$ СВР-расщепления в ФЭ СВР, как следует из результатов реинтерпретации результатов работы [15]. В то же время для ФЭ СВР в карбиноидах данное предположение не выполняется, о чем свидетельствуют ширины мод ФЭ СВР, представленных на рис.1, рис.2. Интересным представляется также и то, что только две колебательные моды оптически активных центров в физически совершенно различных системах - карбиноидах и фуллеридах - участвуют в формировании ФЭ СВР. Следовательно, резонно предполагать, что в исследованных авторами [15] монокристаллах фуллерида как и в карбиноидах реализуется формирование двух (квази)одномерных ферроэлектрически упорядоченных структур, в котором принимают участие две неэквивалентные π−подсистемы.

**Выводы**

Таким образом, идентифицировано новое физическое явление - ферроэлектрический спин-волновой резонанс (ФЭ СВР), состоящее в характерном расщеплении линий колебательных (электронно-колебательных) уровней в оптических спектрах взаимодействующих локализованных центров (на примере системы спин-Пайерлс солитонов в карбиноидах). Положение линий в спектре ФЭ СВР в линейном приближении определяется квадратичным законом дисперсии. В данном приближении установлено теоретически и подтверждено экспериментально соотношение между амплитудами резонансных мод – амплитуда моды обратно пропорциональна номеру моды. Подтвержден экспериментально вывод работы [1] об удвоении параметра спин-волнового расщепления при КР-детектировании ФЭ СВР, являющийся независимым аргументом для его идентификации. Установлено, что исследуемые системы могут проявлять одновременно как ферроэлектрические, так и ферромагнитные свойства. ФЭ ЭСВР является первым из стационарных оптических аналогов явлений и эффектов, сопутствующих магнитному резонансу. Предложена новая интерпретация экспериментальных результатов работы [15] как ФЭ СВР, означающего, в свою очередь, что в исследованных авторами [15] монокристаллах фуллерида реализуется одномерное ферроэлектрическое упорядочение, хотя и весьма слабое, энергия которого почти в 40 раз меньше энергии ферроэлектрического упорядочения в карбиноидах.

**Благодарности**

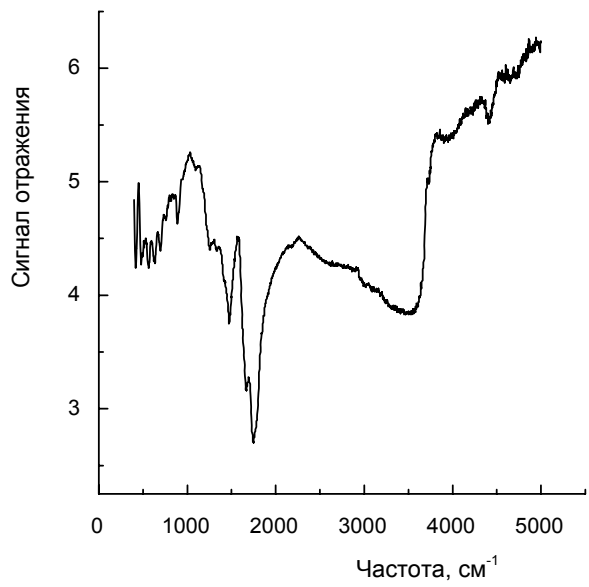
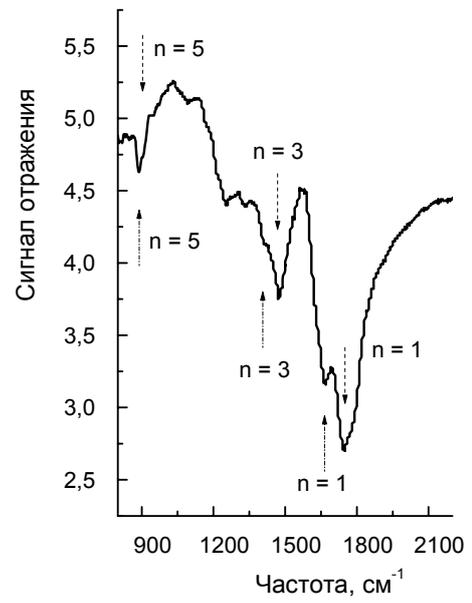

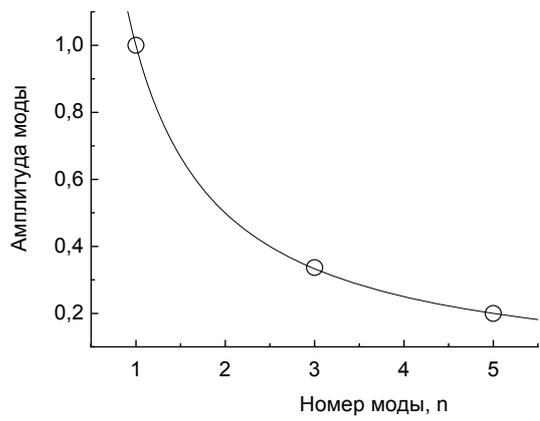

Рис.1

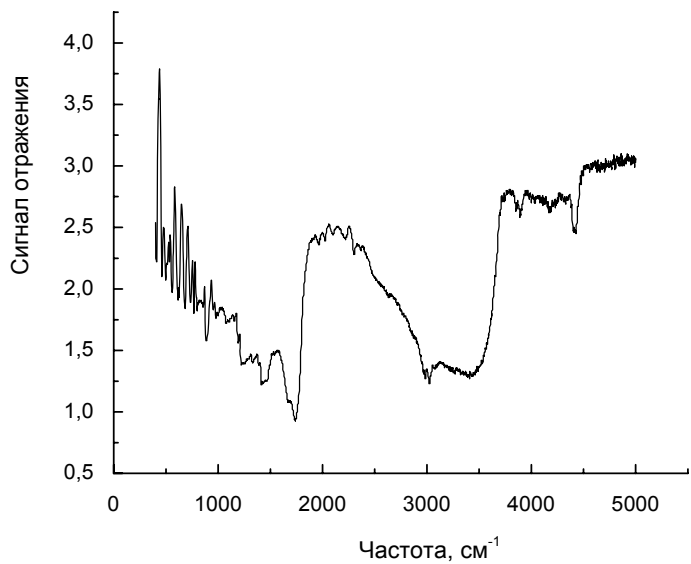
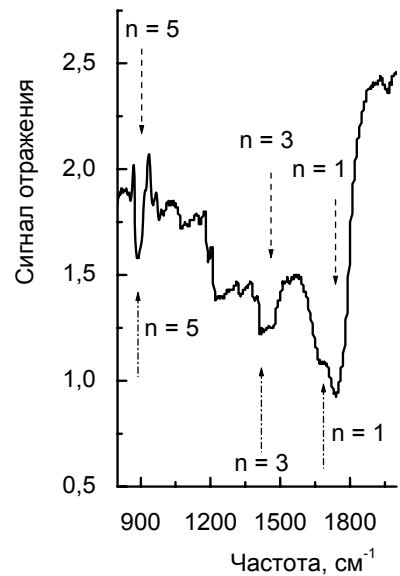

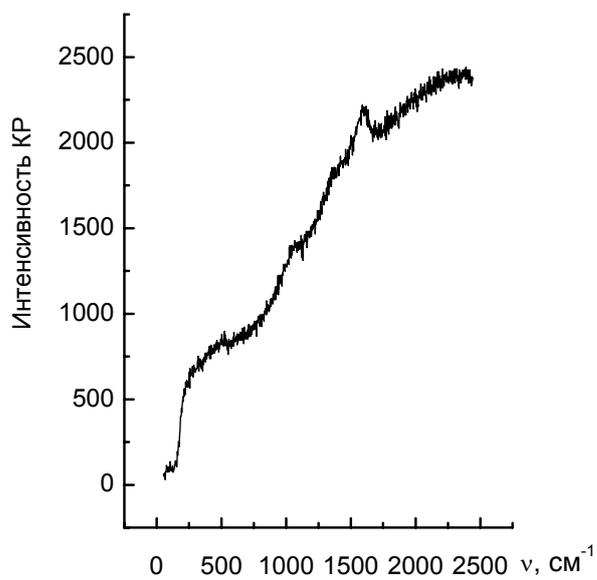
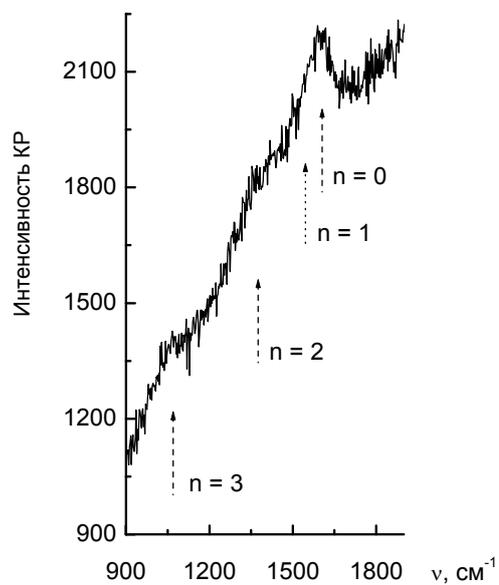

Рис.2

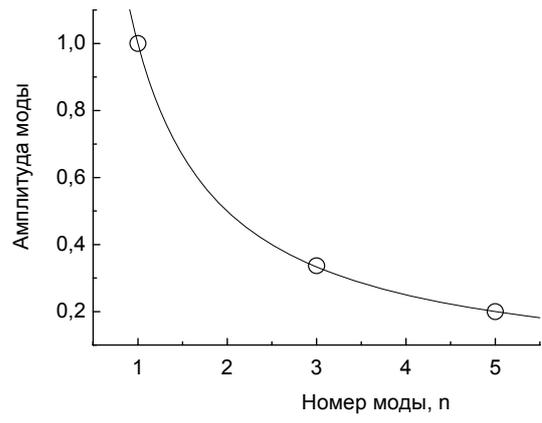

Рис.3

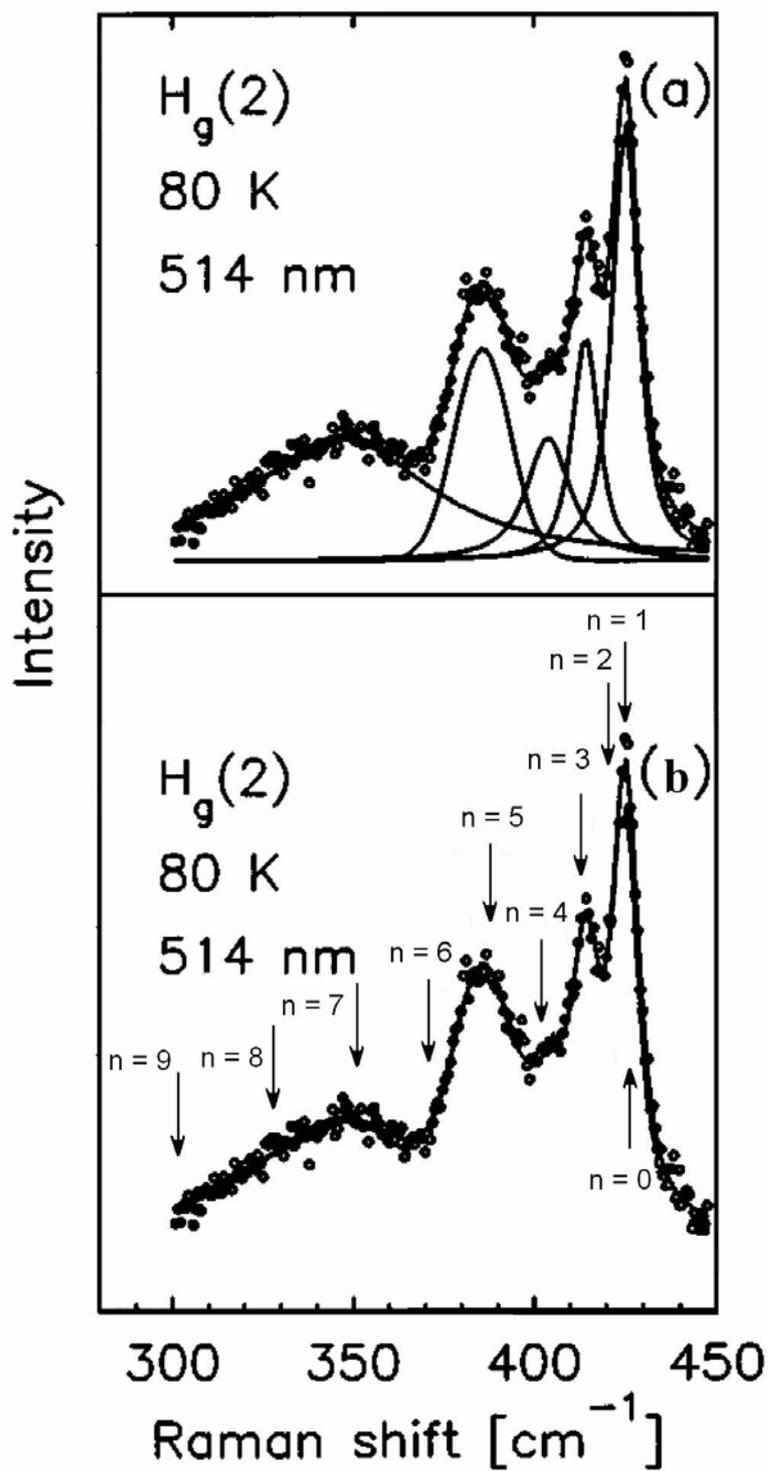

Рис.4

Подписи к рисункам

Рис.1. Спектральное распределение интенсивности сигнала ИК-отражения для *A*-образца карбиноидной пленки, (а) - общий вид спектрального распределения, (б) – детальный вид спектрального распределения в области 800 - 2200 см$^{-1}$.

Рис.2. Спектральное распределение интенсивности сигнала ИК-отражения и КР-сигнала для *B*-образца карбиноидной пленки, (*а*) - общий вид сигнала ИК-отражения, (*б*) – детальный вид сигнала ИК-отражения в области 800 – 2200 см$^{-1}$, (в) – общий вид КР-сигнала, (г) – детальный вид КР-сигнала в области 900 – 1900 см$^{-1}$.

Рис.3. Зависимость амплитуд мод в $\{a_n\}_{[A]}$-наборе для *A*-образца при резонансе от номера моды. Кружками обозначены экспериментальные значения. Сплошная линия – теоретическая зависимость согласно (9).

Рис.4а. Детальный вид спектрального распределения КР-сигнала *Hg*(2) моды в монокристаллах $K_3C_{60}$ при 80 K, воспроизведенный из [15],

(b) спектральное распределение, представленное на рис.4а и интерпретируемое как ФЭ СВР.

# FERROELECTRICAL SPIN WAVE RESONANCE

*YEARCHUCK D.P, YERCHAK E.D, KIRILENKO A.I, POPECHITS V.I*

dpy@tut.by

**Summary**

New phenomenon is experimentally identified: ferroelectrical spin wave resonance (FE SWR), which consist in characteristic splitting of vibration (electronic-vibration) levels in optical spectra of interacting localized centers. Spectral positions of ESWR lines are determined in linear approach by quadratic dispersion law. It has been found that the values of resonance mode amplitudes are inversely proportional to mode numbers (by low excitation level). The prediction that Raman-ESWR and IR-ESWR are characterized by splitting constants with different values has been confirmed. Their ratio is approximately equal two (by the frequencies of zero modes reduced to the same value). It is independent argument for FE SWR identification.

# ФЕРРОЭЛЕКТРИЧЕСКИЙ СПИН-ВОЛНОВОЙ РЕЗОНАНС

ЕРЧАК Д.П [a], ЕРЧАК Е.Д [b], КИРИЛЕНКО А.И [a], ПОПЕЧИЦ В.И [a,b]

Идентифицировано явление ферроэлектрического спин-волнового резонанса (ФЭ СВР), состоящее в характерном расщеплении линий колебательных (электронно-колебательных) уровней в оптических спектрах взаимодействующих локализованных центров (на примере системы спин-Пайерлс солитонов в карбиноидах). Положение линий в спектре ФЭ СВР в линейном приближении определяется квадратичным законом дисперсии. В линейном приближении установлено теоретически и подтверждено экспериментально соотношение между амплитудами резонансных мод – амплитуда моды обратно пропорциональна номеру моды. Подтверждено экспериментально предсказание об удвоении параметра спин-волнового расщепления в КР-спектре ФЭ СВР в сравнении с его величиной в спектре ФЭ СВР, формирующемся при ИК-отражении, что является независимым аргументом для идентификации ФЭ СВР.
Ил.4, Библиогр. – 17 назв.